\documentclass[final,5p,times,twocolumn]{elsarticle}
\usepackage{graphics}
\usepackage{amsmath}
\biboptions{square,numbers,sort&compress}

\usepackage{amssymb}

\journal{Journal of Magnetic Resonance}

\begin{document}

\begin{frontmatter}

\title{Experimental implementation of quantum information processing by Zeeman perturbed nuclear quadrupole resonance}

\author[ufscar]{J. Teles\corref{cor1}}
\ead{jocoteles@gmail.com}

\author[ifsc]{C. Rivera-Ascona}

\author[ifsc]{R. S. Polli}

\author[ifsc]{R. Oliveira-Silva}

\author[ifsc]{E. L. G. Vidoto}

\author[ifsc]{J. P. Andreeta}

\author[ifsc]{T. J. Bonagamba}

\cortext[cor1]{Corresponding author}

\address[ufscar]{Departamento de Ci\^{e}ncias da Natureza, Matem\'{a}tica e Educa\c{c}\~{a}o, Universidade Federal de S\~{a}o Carlos, Caixa Postal 153, 13600-970, Araras, SP, Brazil}

\address[ifsc]{Instituto de F\'{i}sica de S\~{a}o Carlos, Universidade de S\~{a}o Paulo, Caixa Postal 369, 13560-970, S\~{a}o Carlos, SP, Brazil}

\begin{abstract}
Nuclear magnetic resonance (NMR) has been widely used in the context of quantum information processing (QIP). However, despite the great similarities between NMR and nuclear quadrupole resonance (NQR), no experimental implementation for QIP using NQR has been reported. We describe the implementation of basic quantum gates and their applications on the creation of pseudopure states using linearly polarized radiofrequency pulses under static magnetic field perturbation. The NQR quantum operations were implemented using a single crystal sample of KClO$_{3}$ and observing $^{35}$Cl nuclei, which posses spin 3/2 and give rise to a 2-qubit system. The results are very promising and indicate that NQR can be successfully used for performing fundamental experiments in QIP. One advantage of NQR in comparison to NMR is that the main interaction is internal to the sample, which makes the system more compact, lowering its cost and making it easier to be miniaturized to solid state devices.

\end{abstract}


\end{frontmatter}


\section{Introduction}

Many nuclear magnetic resonance (NMR) systems have been used in quantum information processing (QIP) \cite{Vandersypen2004,Ramanathan2004,Suter2008}. The first proposals were performed using liquid state NMR, followed by solid state experiments. Nano-scale devices and optically detected NMR were also proposed \cite{Fisher2003, Leskowitz2003, Itoh2005, Yusa2005, Baugh2006, Childress2006, Hanson2008}. In all these cases, the Zeeman interaction with an external magnetic field is the main interaction, where the perturbations range from scalar couplings in liquids, direct dipole couplings in solids and quadrupole interaction in solids and liquid crystals.

In nuclear quadrupole resonance (NQR), the main interaction is due to the nuclear electric quadrupole moment in the presence of a local electric field gradient. As in the NMR case, the sole main interaction does not allow the general transitions between all logic states. The first theoretical proposal of the use of pure nuclear quadrupole resonance in the context of QIP was done by Furman and Goren \cite{Furman2002}, where the use of two linearly polarized radiofrequency (RF) fields at arbitrary relative angles are responsible for the degeneracy lifting of the four magnetic states in the case of a spin 3/2, while a third RF field or the amplitude modulation of one of the two former fields promote energy transitions. Possa et al \cite{Possa2011}, proposed a simulation of the NMR-NQR system in an arbitrary configuration of elliptically polarized RF fields and quadrupole and Zeeman interactions. Particularly, aiming QIP applications, they showed how to prepare pseudo-pure states and how to implement basic quantum gates in pure NQR using circularly polarized RF fields and double-quantum excitation.

Here we describe the first NQR experimental protocol to implement QIP operations in a two q-bit spin 3/2 system using linearly polarized radiofrequency pulses under static magnetic field perturbation. This protocol was implemented in the spin 3/2 $^{35}$Cl nuclei of a single crystal sample of KClO$_3$ \cite{Zeldes1957}. As result, we obtained all the four pseudo-pure states associated with the computational basis and applied on them the Controlled-not (CNOT) and Hadamard gates. The reading of the resulting states were performed by approximately $\pi/2$ RF pulses.

One of the advantages of the use of NQR systems for QIP resides in the compactness of the experimental system, since the main interaction is produced internally by the own sample, making unnecessary the use of high magnetic fields normally produced by superconductor magnets, as in the case of NMR. In the case of NQR, the perturbative magnetic field can be generated by a small Helmholtz coil pair, and represents little cost to the system. Moreover, under Zeeman perturbation we avoid the use of crossed RF coils necessary for distinguishing different transitions in the pure NQR case, by the use of circularized RF fields.

\section{Theory}

Here we will describe the form of the main operators in the quadrupole interaction picture using the secular approximation in the simplified case of a cylindrical symmetric electric field gradient. The expressions for the energy levels and for the RF pulse operators will be applied in section \ref{Experimental} in order to implement the quantum gates optimizations and the experimental set up.

\subsection{The Zeeman perturbed NQR Hamiltonian}

We will consider the case in which the main contribution to the nuclear Hamiltonian is due to the interaction of the nuclear electric quadrupole moment with the local electric field gradient (EFG), where this gradient is assumed to be cylindrically symmetric. The symmetry axis of the EFG tensor will be designated by the vector $\vec{G}$ and will be oriented along the $z$ direction, as usual. We will consider the case of a single-crystalline sample having only one EFG symmetry axis direction per unity cell. To the main term will be added an external static magnetic field $\vec{B}_{0}$, oriented in an arbitrary direction constrained to the $xz$ plane and making an angle $\theta$ with the $z$-axis. The time dependent perturbation necessary for the quantum transitions (quantum gates) will be accomplished by a linearly polarized oscillating magnetic field $\vec{B}_{1}$ (RF pulses) constrained to the $xy$ plane and making an angle $\phi$ with the $x$-axis. Fig.~1 illustrates these choices. 

\begin{figure}[h!]
\centering
\includegraphics[scale=1.0]{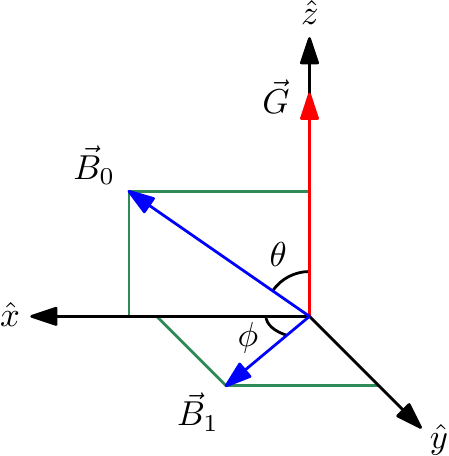}
\caption{Definition of the angles and orientations of the main interactions in the Zeeman perturbed NQR system.}
\end{figure}

Under those conditions, the NQR Hamiltonian is written in the laboratory frame as:
\begin{align}\label{eq1}
H(t)&=H_{0}+H_{RF}(t)\;.
\end{align}

The RF unperturbed term, $H_{0}=H_{Q}+H_{Z}$, contains the contribution of the quadrupole interaction and the Zeeman perturbation with the static magnetic field, while the $H_{RF}$ term is due to the RF field.

The quantum operators are given by:
\begin{equation}
H_{Q}=\hbar\frac{\omega_{Q}}{2}\left[I_{z}^{2}-\hat{1}I(I+1)/3\right]\;,
\end{equation}
\begin{equation}\label{eq1p4}
H_{Z}=-\hbar\omega_{0}(I_{x}\sin\theta+I_{z}\cos\theta)\;\mathrm{, and}
\end{equation}
\begin{equation}\label{eq1p5}
H_{RF}(t)=-2\hbar\omega_{1}\cos(\omega t+\alpha)e^{-i\phi I_{z}}I_{x}e^{i\phi I_{z}}\;,
\end{equation}
where $I$ is the spin quantum number, $I_{x}$, $I_{y}$, and $I_{z}$ are the dimensionless Cartesian angular operator components, $\omega_{0}=\gamma B_{0}$ and $\omega_{1}=\gamma B_{1}$ are the Zeeman couplings with the static and RF fields of amplitudes $B_{0}$ and $B_{1}$, respectively. The RF field is applied with frequency $\omega$ and initial phase $\alpha$.

Since the Zeeman term is a perturbation of the main Hamiltonian, the quadrupole coupling $\omega_{Q}$ is much greater than $\omega_{0}$. In the absence of the $B_{0}$ field, the quantum states are two-fold degenerate, and the quantum transitions occur at frequencies multiple of $\omega_{Q}$, accordingly with Fig.~2.

\begin{figure}[h!]
\centering
\includegraphics[scale=0.6]{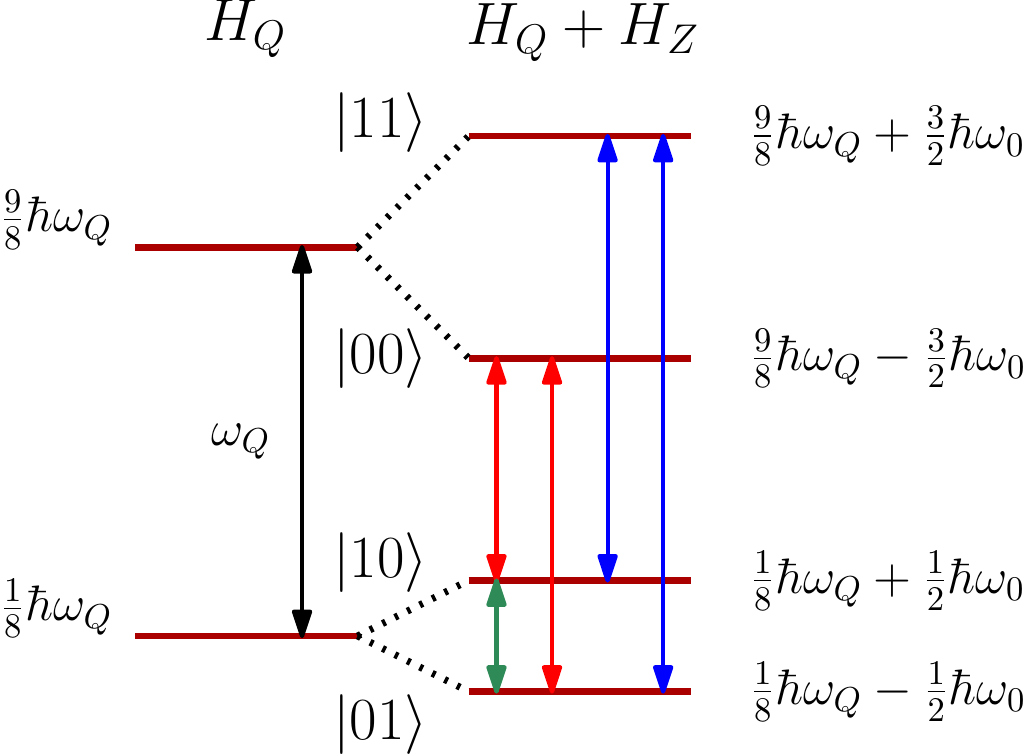}
\caption{Energy states and the allowed transitions in pure NQR ($H_Q$) and Zeeman perturbed NQR ($H_{Q}+H_{Z}$) for the $\theta=0$ case. The QIP logical labelling is presented. Transitions between states $|01\rangle$ and $|10\rangle$ are ignored as they are far away from $\omega_{Q}$.}
\end{figure}

\subsection{Quadrupole interaction picture}

To facilitate the solution of the Schr\"{o}dinger equation for the nuclear state $|\psi\rangle$ we will apply the following unitary operation on going to the quadrupole interaction picture:
\begin{equation}
U(t)=e^{\frac{i}{2}\omega I_{z}^{2}t}\;.
\end{equation}

The transformed Hamiltonian is given by:
\begin{equation}
\tilde{H}=UHU^{-1}-i\hbar U\frac{d}{dt}U^{-1}\;.
\end{equation}

Applying the above transformation in Eq.~(\ref{eq1}) results in:
\begin{equation}\label{eq2}
\tilde{H}(t)=\frac{\hbar}{2}\Delta\omega_{Q}I_{z}^{2}+\tilde{H}_{Z}
+\tilde{H}_{RF}(t)\;,
\end{equation}
where $\Delta\omega_{Q}=\omega_{Q}-\omega$ and we neglected the term proportional to identity. All operators in the quadrupole picture are represented by the tilde symbol. The transformed operators in Eq.~(\ref{eq2}), are calculated by:
\begin{equation}
U\frac{d}{dt}U^{-1}=-\frac{i}{2}\omega I_{z}^{2}\;,
\end{equation}
\begin{equation}
UI_{z}U^{-1}=I_{z}\;,
\end{equation}
\begin{equation}
UI_{z}^{2}U^{-1}=I_{z}^{2}\;\mathrm{, and}
\end{equation}
\begin{equation}
\left[UI_{x}U^{-1}\right]_{i,i\pm1}=\exp\left[\frac{i}{2}\omega t\left(m_{i}^{2}-m_{i\pm1}^2\right)\right][I_{x}]_{i,i\pm1}\;.
\end{equation}

For the $I_x$ operator, only the non-null matrix elements in the $I_z$ eigenstates basis were shown.

The RF field will induce transitions between the quantum states for values of $\omega$ close to $\omega_{Q}$. In this condition, we can neglect, in first order approximation, any terms that oscillate in Eq.~(\ref{eq2}) with frequencies in the order of $\omega$, since these terms have frequencies much higher than $\Delta\omega_{Q}$ and $\omega_{0}$. In evaluating the matrix elements of $\tilde{H}_{Z}$, we see that only their central transition elements $(\pm1/2,\mp1/2)$ are independent of $\omega$, making these elements the only ones that must be considered. Otherwise, the matrix elements of $\tilde{H}_{RF}$ are multiplied by the function $\cos(\omega t+\alpha)$ in Eq.~(\ref{eq1p5}). In this case, the oscillating terms of the satellite transitions $(\pm3/2,\mp1/2)$ are cancelled by the cosine terms. The same does not happen with the central transition elements, which are the only ones to be neglected in the $\tilde{H}_{RF}$ operator.

To illustrate the above discussion, let us calculate the matrix elements for the specific case of a spin~3/2 nucleus in the $I_z$ eigenstates basis:
\begin{align}
\tilde{I}_{x}&=UI_{x}U^{-1}\nonumber \\
&=\frac{1}{2}
\begin{bmatrix}
	0 & \sqrt{3}e^{i\omega t} & 0 & 0 \\
	\sqrt{3}e^{-i\omega t} & 0 & 2 & 0 \\
	0 & 2 & 0 & \sqrt{3}e^{-i\omega t} \\
	0 & 0 & \sqrt{3}e^{i\omega t} & 0
	\end{bmatrix}\;,
\end{align}
which in first order approximation reduces to:
\begin{equation}\label{eq2p5}
\tilde{I}_{x}\approx\begin{bmatrix}
	0 & 0 & 0 & 0 \\
	0 & 0 & 1 & 0 \\
	0 & 1 & 0 & 0 \\
	0 & 0 & 0 & 0
\end{bmatrix}\;.
\end{equation}

The same approximation applied to the RF Hamiltonian results in:
\begin{align}\label{eq3}
\frac{\tilde{H}_{RF}}{\hbar\omega_{1}}&=-2\cos(\omega t+\alpha)U(I_{x}\cos\phi+I_{y}\sin\phi)U^{-1}\nonumber \\
&\approx -\frac{\sqrt{3}}{2}
\begin{bmatrix}
	0 & e^{-i\phi_{+}} & 0 & 0 \\
	e^{i\phi_{+}} & 0 & 0 & 0 \\
	0 & 0 & 0 & e^{-i\phi_{-}} \\
	0 & 0 & e^{i\phi_{-}} & 0
	\end{bmatrix}\;,
\end{align}
where $\phi_{\pm}\equiv\phi\pm\alpha$. Therefore, for $B_{1}$ time independent, we obtained the time independent effective Hamiltonian $\tilde{H}$ in first order approximation.

\subsection{Subspace diagonalization}

Observing the representation of the operator $\tilde{I}_{RF}$ in Eq.~(\ref{eq3}), we see that there are non-null elements only in the individual subspaces of each q-bit. That could be a problem to produce general two q-bit logic gates. However, the existence of the $\tilde{I}_{x}$ operator due to the static field perturbation, makes possible the q-bits conditional evolution, as can be seen in the $(\pm1/2,\mp1/2)$ subspace in Eq.~(\ref{eq2p5}). In order to obtain the $\tilde{H}_{0}$ eigenvalues, it is necessary diagonalize the RF unperturbed Hamiltonian. The operator $R$ that diagonalizes $\tilde{H}_{0}$ in the case of a spin 3/2 subspace, is given by:
\begin{equation}\label{eq4}
	R=\begin{bmatrix}
			1 & 0 & 0 & 0 \\
			0 & f_{+} & f_{-} & 0 \\
			0 & f_{-} & -f_{+} & 0 \\
			0 & 0 & 0 & 1
	\end{bmatrix}\;,
\end{equation}
where:
\begin{equation}
	f_{\pm}=\sqrt{\frac{1}{2}\pm\frac{1}{2\sqrt{1+4\tan^{2}\theta}}}\;.
\end{equation}

For the spin 3/2, the eigenvalues of the RF unperturbed Hamiltonian obtained with the $R$ transformation are:
\begin{equation}\label{eq5}
E_{(+3/2)_R}=\frac{\hbar}{8}(9\Delta\omega_{Q}-12\omega_{0}\cos\theta)\;,
\end{equation}
\begin{equation}
E_{(+1/2)_R}=\frac{\hbar}{8}(\Delta\omega_{Q}-4\omega_{0}g\cos\theta)\;,
\end{equation}
\begin{equation}
E_{(-1/2)_R}=\frac{\hbar}{8}(\Delta\omega_{Q}+4\omega_{0}g\cos\theta)\;\mathrm{, and}
\end{equation}
\begin{equation}\label{eq5p4}
E_{(-3/2)_R}=\frac{\hbar}{8}(9\Delta\omega_{Q}+12\omega_{0}\cos\theta)\;,
\end{equation}
where $g=\sqrt{1+4\tan^{2}\theta}$. The RF operator, in the new $R$ basis, is given by:
\begin{equation}\label{eq5p5}
\frac{\hat{H}_{RF}}{\hbar\omega_{1}}=-\frac{\sqrt{3}}{2}
\begin{bmatrix}
	0 & e^{-i\phi_{+}}f_{+} & e^{-i\phi_{+}}f_{-} & 0 \\
	e^{i\phi_{+}}f_{+} & 0 & 0 & e^{-i\phi_{-}}f_{-} \\
	e^{i\phi_{+}}f_{-} & 0 & 0 & -e^{-i\phi_{-}}f_{+} \\
	0 & e^{i\phi_{-}}f_{-} & -e^{i\phi_{-}}f_{+} & 0
	\end{bmatrix}\;.
\end{equation}

All operators in the quadrupole picture and transformed by $R$ are represented by the hat symbol.

\subsection{State evolution and measurement}

The Zeeman perturbed NQR signal can be calculated by the following trace equation:
\begin{equation}\label{eq6}
S(t)\propto\mathrm{Tr}\left\{\hat{\rho}\cdot \hat{U}_{0}^{-1}(t)\hat{H}_{RF}\hat{U}_{0}(t)\right\}\;,
\end{equation}
where $\hat{U}_{0}(t)=e^{-\frac{i}{\hbar}\hat{H}_{0}t}$ is the evolution operator during the reading interval in the laboratory frame, that is, considering $\Delta\omega_{Q}=\omega_{Q}$ in Eqs.~(\ref{eq5}) to (\ref{eq5p4}). The resulting density matrix $\hat{\rho}=\hat{P}\rho_{eq}\hat{P}^ {-1}$ is obtained by the application of the propagator:
\begin{equation}\label{eq6p1}
\hat{P}=\prod_{n}e^{-\frac{i}{\hbar}\hat{H}_{n}\Delta t_{n}}\;,
\end{equation}
where $\Delta t_{n}$ is sufficiently small such that $\hat{H}_{n}$ can be considered time independent in the corresponding interval. It is possible to factorize the operator $e^{-i\phi(I_{z_1}+I_{z_2})}$ from $\hat{H}_{RF}$ and $\hat{H}_{0}$, where $I_{z_1}$ and $I_{z_2}$ are spin 1/2 operators, such that, the signal $S(t)$ is independent of the polarization angle $\phi$.

The initial equilibrium state $\rho_{eq}$ is given by the Boltzmann statistics in the high temperature approximation:
\begin{equation}\label{eq6p2}
\rho_{eq}=\frac{1}{Z}\left(\hat{1}-\frac{H_{0}}{kT}\right)\;,
\end{equation}
where $Z$ is the partition function, $k$ the Boltzmann constant, and $T$ the sample temperature.

Applying Eqs. (\ref{eq5}) to (\ref{eq5p5}) into Eq. (\ref{eq6}) results in the following signal expression for spin 3/2:
\begin{align}\label{eq7}
S(t)\propto&\;\hat{\rho}^{*}_{1,2}f_{+}e^{i\nu_{1,2}t}
+\hat{\rho}^{*}_{1,3}f_{-}e^{i\nu_{1,3}t}+\nonumber\\
&+\hat{\rho}_{2,4}f_{-}e^{-i\nu_{2,4}t}
-\hat{\rho}_{3,4}f_{+}e^{-i\nu_{3,4}t}\;,
\end{align}
where the indexes $(1,2,3,4)$ correspond, respectively, to the quantum numbers $(\frac{3}{2},\frac{1}{2},-\frac{1}{2},-\frac{3}{2})_{R}$, and the Bohr frequencies $\nu_{i,j}=(E_{i}-E_{j})/\hbar$ are obtained from Eqs. (\ref{eq5}) to (\ref{eq5p4}).

Therefore, in the presence of the static magnetic field perturbation, the two degenerate states of the spin 3/2 are lifted, producing up to four distinct energy levels and up to four observable transitions, as obtained in Eq.~(\ref{eq7}) and illustrated by Fig.~2. The density matrix elements can also be computed by Eq.~(\ref{eq7}), where reading pulses can be applied prior to acquisition in order to transfer the various coherences of $\hat{\rho}$ to the coherences indicated in Eq.~(\ref{eq7}) -- a process known as Quantum State Tomography in QIP terminology, procedure which was not developed in this work, but it is under development in our group.

\section{\label{Experimental} Experimental}

\subsection{KClO$_3$ Single-crystal}

The $^{35}$Cl nuclei posses spin 3/2 with a gyromagnetic ratio of 4.176~MHz/T, and in a single crystal of KClO$_3$ presents a quadrupole coupling $\omega_{Q}/2\pi=28.1$~MHz. There is only one EFG symmetry axis per unit cell, since all molecules are crystallographic equivalent \cite{Zeldes1957}. 

The single crystals of the potassium chlorate were grown from aqueous solution method. A potassium chlorate solution was prepared by dissolving the Merck GR (for analysis) KClO$_3$ powder in distilled water. The solution, after being filtered at temperature of  35~degree Celsius, was sealed  (to prevent solvent loss by evaporation) and placed in a heat bath. The crystal seeds were nucleated using the slow cooling process up to 30~$^\circ$C with cooling rate of 0.5-0.3~$^\circ$C/day. Although irregular growths were observed in some of the experiments, with tendencies to grow dendrites, needle formed crystals and rough surfaces, transparent and high optical quality potassium chlorate single crystals were grown, using slow temperature reduction -- 0.1~$^\circ$C/day -- on a rotating seed holder at 15~rpm. The whole preparation process, the stable growth conditions as well the growth habit control will be described elsewhere.

\subsection{Crystal orientation and the choice of $\theta$}

The Zeeman perturbed NQR Hamiltonian and, consequently, the quantum states evolution, depend on the $\theta$ angle between the static magnetic field $\vec{B}_{0}$ and the direction of the principal axis of the EFG tensor, $\vec{G}$. In order to easily implement any $\theta$ angle, we built a RF probe inside the goniometer illustrated at Fig.~3. In this system, the crystalline sample is kept fixed inside the RF coil and both stay fixed inside the probe holder box. The latter, in turn, can be rotated around the vertical $\hat{v}$ and the horizontal $\hat{h}$ axes. The field $\vec{B}_{0}$ is stationary and coplanar with the horizontal plane.

\begin{figure}[h!]
\centering
\includegraphics[scale=1.0]{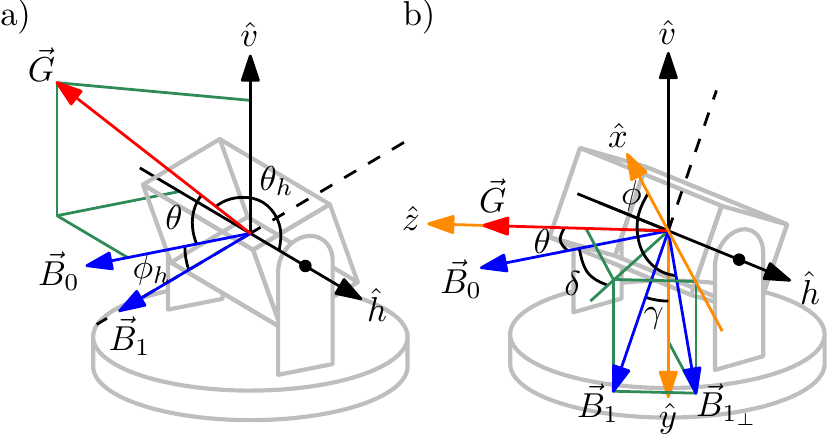}
\caption{a) Initial configuration of the goniometer, where the spectral line frequencies are measured as function of $\phi_{h}$. b) The EFG vector $\vec{G}$ is brought to the horizontal plane, which also contains the vectors $\vec{B_{0}}$, $\hat{x}$, $\hat{z}$, and $\hat{h}$. In this last configuration, any $\theta$ angle can be obtained just by rotations around $\hat{v}$.}
\end{figure}

The crystal was positioned with its longer dimension aligned with the RF coil axis in order to have a better filling factor, which was even improved using a solenoid of elliptical cross section. The drawback with this squeezed coil design was a decrease in the $B_1$ amplitude homogeneity. With those choices, $\vec{G}$ points at a direction that, in general, does not belong to the horizontal plane. Fig.~3a illustrates that situation, where $\vec{G}$ makes an initial $\theta$ angle with $\vec{B}_{0}$. The first procedure is, therefore, to bring $\vec{G}$ to the horizontal plane. This can be done by setting $\hat{h}$ perpendicular to $\vec{B}_{0}$ and observing the frequencies of the spectral lines as function of the horizontal rotation angle $\phi_{h}$. It can be shown that the $\theta$ dependency in Eqs. (\ref{eq5}) to (\ref{eq5p4}) is:
\begin{equation}\label{eq8}
\cos\theta=\sin\theta_{h}\cos\phi_{h}\;,
\end{equation}
where $\theta_{h}$ is the angle between $\vec{G}$ and the rotation axis $\hat{h}$. With the aid of Eq.~(\ref{eq8}) it can be shown that the maxima and minima for all frequency functions $\nu_{i,j}(\phi_{h})$ occur at $\phi_h=n\pi$ for $n$ integer -- angles at which $\vec{G}$ belongs to the horizontal plane. Fig.~4 shows the experimental frequencies of the $^{35}$Cl spectral line frequencies as function of $\theta$. From this data, it was possible to choose any $\theta$ angle by just rotating the vertical axis of the goniometer in the condition illustrated in Fig.~3.b.

\begin{figure}[h!]
\centering
\includegraphics[scale=0.9]{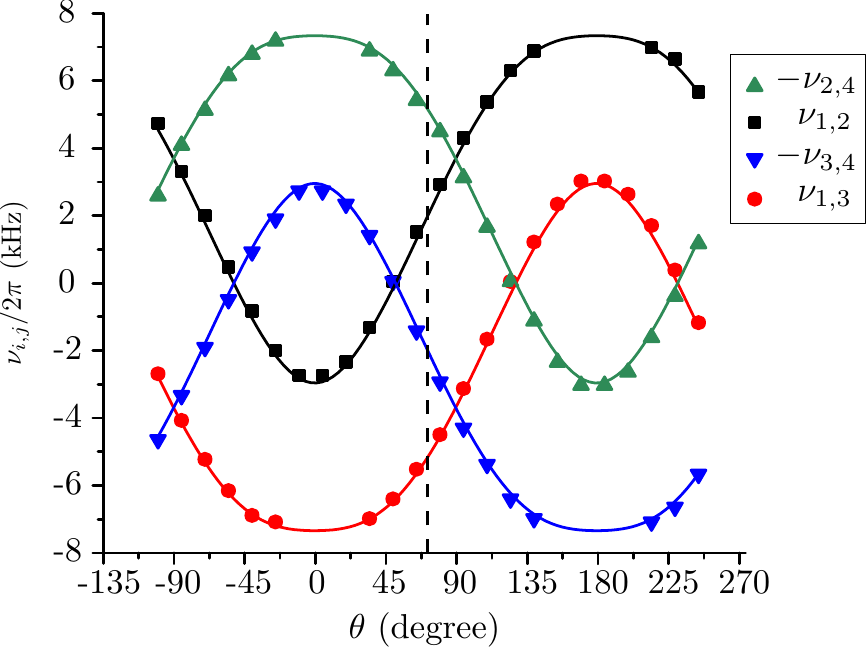}
\caption{Experimental data of the $^{35}$Cl spectral line frequencies $\nu_{i,j}$ as function of the $\theta$ angle between $\vec{G}$ and $\vec{B}_{0}$. Error bars are not represented, since they are smaller then the data symbols. The continuous lines are the theoretical $\nu_{i,j}$ functions fitted to the experimental data. The vertical dashed line corresponds to the angle $\theta=71.3^{\circ}$ used in this work to implement QIP.}
\end{figure}

Since just the perpendicular component of $\vec{B}_{1}$ to $\vec{G}$ is effective in inducing transitions, we obtained the following expression:
\begin{equation}\label{eq10}
B_{1_\perp}=B_{1}\sqrt{1+
\sin^{2}\gamma\left[\sin^{2}(\theta+\delta)-1\right]}\;,
\end{equation}
where $\gamma$ is the angle that $\vec{B}_{1}$ does with $\hat{v}$ and $\delta$ is the angle that the $\vec{B}_{1}$ projection onto the $xz$ plane does with $\vec{B}_{0}$. This geometry is indicated in Fig.~3b.

We envisage at least three criteria for the choice of the $\theta$ angle:
\begin{enumerate}
\item Maximize $B_{1_{\perp}}$ in order to apply shorter pulses and, then, minimize relaxation effects. Accordingly to Eq.~(\ref{eq10}), this condition is met for $\theta=\pi/2-\delta$.
\item Choose angles were the frequency functions $\nu_{i,j}(\theta)$ are most sensitive to $\theta$ in order to minimize errors in its determination. From Fig.~4b we can see that this condition is fairly satisfied for the intervals $75^{\circ}\leq\theta\leq105^{\circ}$ and $255^{\circ}\leq\theta\leq285^{\circ}$.
\item Make the spectral lines evenly spaced to one another, such that single transitions can be individually excited with shorter RF pulses and, again, minimize relaxation. This condition corresponds to $\cos^{2}\theta=4/39$. In fact, the greatest separation between spectral lines would be for $\theta=n\pi$ with $n$ integer. However, there would be no contribution of the $I_{x}$ term in Eq.~(\ref{eq1p4}), preventing two q-bit gates implementation.
\end{enumerate}

Since all the three criteria can not be simultaneously satisfied, we have chosen the third one. That corresponds to $\theta\approx 71.3^{\circ}$, which is near the most sensitive region settled by criterion two. Moreover, in our experimental setup we found $\gamma=65^{\circ}$ and $\delta=70^{\circ}$, which gives a suitably high $B_{1_{\perp}}/B_{1}$ ratio of 70.7\%.

\subsection{Relaxation times}

In order to estimate the transverse and longitudinal relaxation times of $^{35}$Cl nuclei in KClO$_3$, we used spin-echo and progressive saturation techniques, respectively.
 
In the case of spin-echo sequence, the experiment was run using 10~$\mu$s and 20~$\mu$s RF pulses for the single-crystal, oriented in such a way that the EFG and the perturbative external magnetic field were aligned, allowing the detection of the echoes intensity, but observing spectra with only two spectral lines. The echo time was changed from 2 to 27~ms in $n$ equally spaced steps of 0.5~ms. The recycle delay was 100~ms. The estimated value for T$_2$ was approximately $(4.6\pm 0.2)$~ms for both lines.

For the progressive saturation experiments, we used only a single 10~$\mu$s RF pulse for the situation where EFG and perturbative external magnetic field were making an angle of 71.3$^\circ$, allowing the observation of the four expected equally spaced peaks in the spectra. The delay times ranged in the interval of 5~ms to 205~ms every 5~ms. The estimated $T_1$ value for all the observed lines was $(32\pm 2)$~ ms. 

\subsection{Quantum gates and pseudo-pure states implementation}

Fig.~5(a) shows the modulus of the $^{35}$Cl spectrum obtained for the equilibrium state after the application of an approximately $\pi/2$ reading pulse, which corresponds to a single RF pulse of 10~$\mu$s. The angle between the static magnetic field and the EFG symmetry axis was kept in $\theta\approx71.3^{\circ}$, giving rise to four equally spaced spectral lines. The frequency separation between adjacent lines was 3.5~kHz, which gives a static magnetic field of 730~$\mu$T. This field was produced by the stray field of a wide horizontal bore superconducting magnet of 2~T. This choice was to take advantage of the laboratory setup, however a small Helmholtz pair could produce a similar magnetic field.

\begin{figure}[h!]
\centering
\includegraphics[scale=0.8]{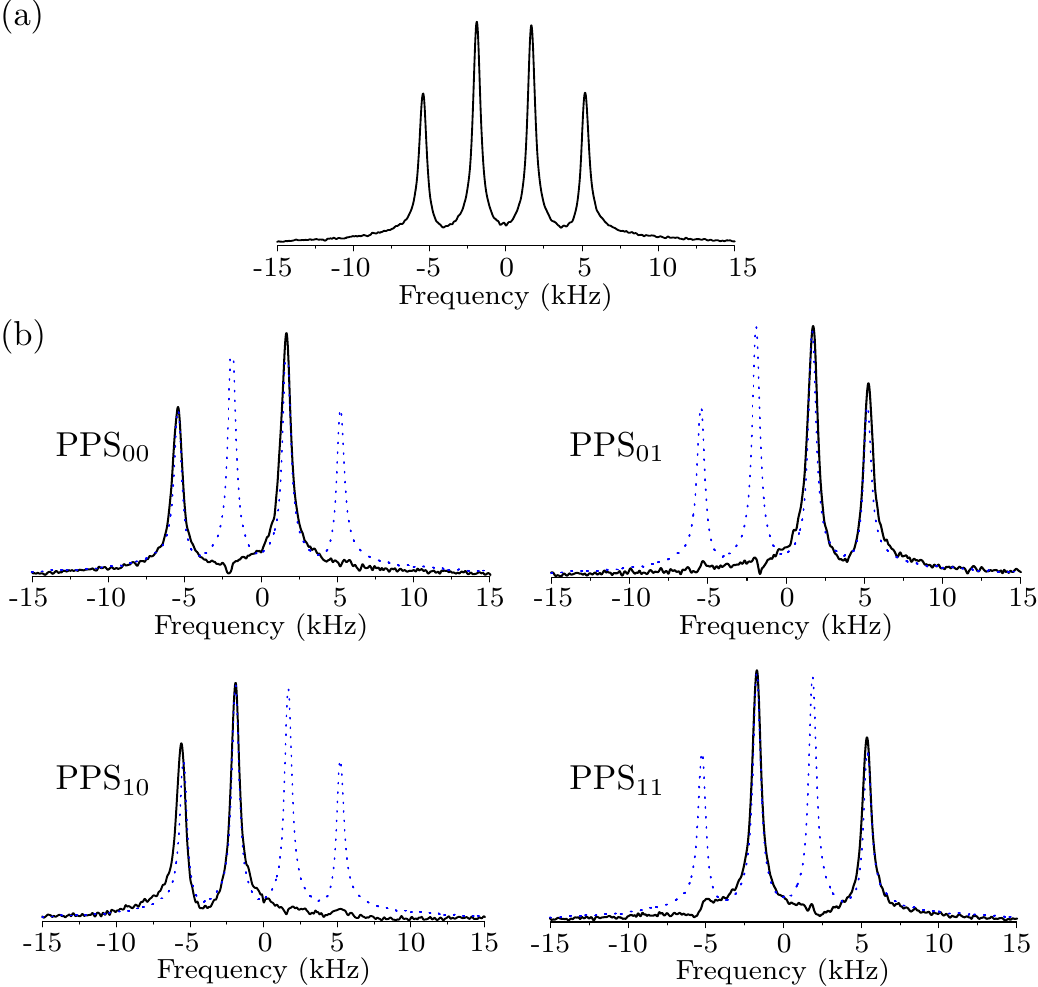}
\caption{$^{35}$Cl spectra obtained after a $\pi/2$ pulse on (a) the equilibrium and (b) the PPS states. The dashed lines correspond to the equilibrium spectrum for comparison.}
\end{figure}

The quantum gates were numerically optimized using Eq.~(\ref{eq6p1}) where the Hilbert-Schmidt inner product was used for fidelity evaluation \cite{Fortunato2002,Teles2007}. Five quantum gates were optimized: (i) Controlled NOT at q-bit $a$ (CNOT$_{a}$), (ii) Controlled NOT at q-bit $b$ (CNOT$_{b}$), (iii) permutation of populations $\frac{3}{2}\leftrightarrow\frac{1}{2}$ (P$_{12}$), (iv) permutation of populations $\frac{3}{2}\leftrightarrow\frac{-1}{2}$ (P$_{13}$), and (v) Hadamard at both q-bits $a$ and $b$ (H$_{ab}$). All gates were implemented with pulses on resonance. The longest gate was the P$_{24}$ whose pulse sequence lasted 230~$\mu$s. Nevertheless, it is approximately one-twentieth of the transverse relaxation time, preventing that irreversible processes took place in the experiments performed in this work.

As in the NMR case, the NQR thermal equilibrium states are far from pure states ($kT\gg\hbar\omega_{Q}$). Nevertheless, the states evolution is almost unitary for times much shorter than the spins relaxation times. These properties characterize thermal NMR and NQR as ensemble quantum computing techniques \cite{Gershenfeld1997,Cory1997}. A common procedure to relate the highly mixed equilibrium states with the pure states used in quantum computing is to apply operators sum on them. It can be shown that the following operators sum:
\begin{equation}\label{eq11p1}
S_{00}=\hat{1}+\mathrm{CNOT}_{a}+\mathrm{CNOT}_{b}\;,
\end{equation}
\begin{equation}\label{eq11p2}
S_{01}=\hat{1}+\mathrm{CNOT}_{a}+\mathrm{P}_{13}\;,
\end{equation}
\begin{equation}\label{eq11p3}
S_{10}=\hat{1}+\mathrm{CNOT}_{b}+\mathrm{P}_{12}\;\mathrm{, and}
\end{equation}
\begin{equation}\label{eq11p4}
S_{11}=\hat{1}+\mathrm{P}_{13}+\mathrm{P}_{12}\;,
\end{equation}
when applied on the deviation equilibrium state:
\begin{equation}
\Delta\rho_{eq}=\frac{\hat{1}}{Z}-\rho_{eq}\propto I_{z}^{2}\;,
\end{equation}
generate the states associated with the computational basis, which are called pseudo-pure states (PPS). To perform the operations of Eqs.~(\ref{eq11p1}) to (\ref{eq11p4}) the pulse sequences corresponding to each quantum gate were implemented one at a time, and the resulting spectra were added. Fig.~6 illustrates the transformation $S_{00}$ on the deviation density matrix populations in order to obtain the pseudo-pure state PPS$_{00}$.

\begin{figure}[h!]
\centering
\includegraphics[scale=1.0]{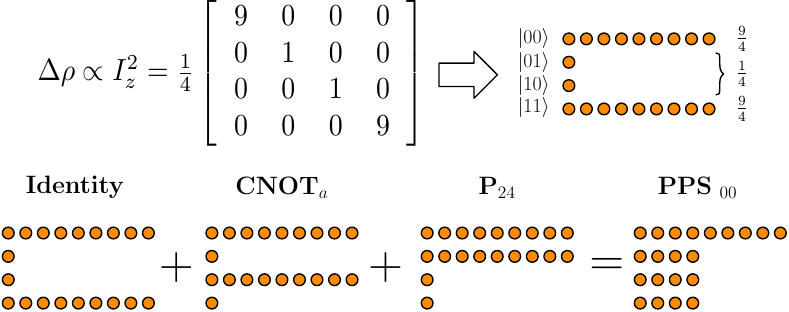}
\caption{Illustration of the procedure for construction of the PPS$_{00}$. The gates CNOT$_a$, P$_{24}$ and Identity (no pulse) are separately applied on the equilibrium state, and the resulting NQR signals are added together.}
\end{figure}

It can be shown that the application of $\pi/2$ reading pulses on each PPS result in the following normalized spectral amplitudes:
\begin{equation}\label{eq34}
\pi/2 \rightarrow \mathrm{PPS}_{00} = 1:0:1:0\;,
\end{equation}
\begin{equation}
\pi/2 \rightarrow \mathrm{PPS}_{01} = 0:0:1:1\;,
\end{equation}
\begin{equation}
\pi/2 \rightarrow \mathrm{PPS}_{10} = 1:1:0:0\;,
\end{equation}
\begin{equation}\label{eq37}
\pi/2 \rightarrow \mathrm{PPS}_{11} = 0:1:0:1\;,
\end{equation}
where each amplitude is normalized relatively to the equilibrium spectrum.

Fig.~5(b) shows the modulus of the spectra obtained for each PPS after the application of the $\pi/2$ reading pulse. We can see a good agreement with the expected amplitudes in Eqs.~(\ref{eq34}) to (\ref{eq37}). The worst result was the PPS$_{01}$ which presented a standard deviation of the expected peak amplitudes relative to the equilibrium spectrum of 6.8\%.

Fig.~7 shows the experimental results of the CNOT$_a$ and CNOT$_b$ gates implementation on each one of the PPS showed in Fig.~5, where the spectra were also obtained after a $\pi/2$ reading pulse. We can see that the spectra are fairly close to the expected states after the application of the CNOT gates. In this case, the worst result was the operation CNOT$_b|01\rangle$ whose spectrum presented a standard deviation of the expected peak amplitudes relative to the equilibrium spectrum of 11\%.

\begin{figure}[h!]
\centering
\includegraphics[scale=0.8]{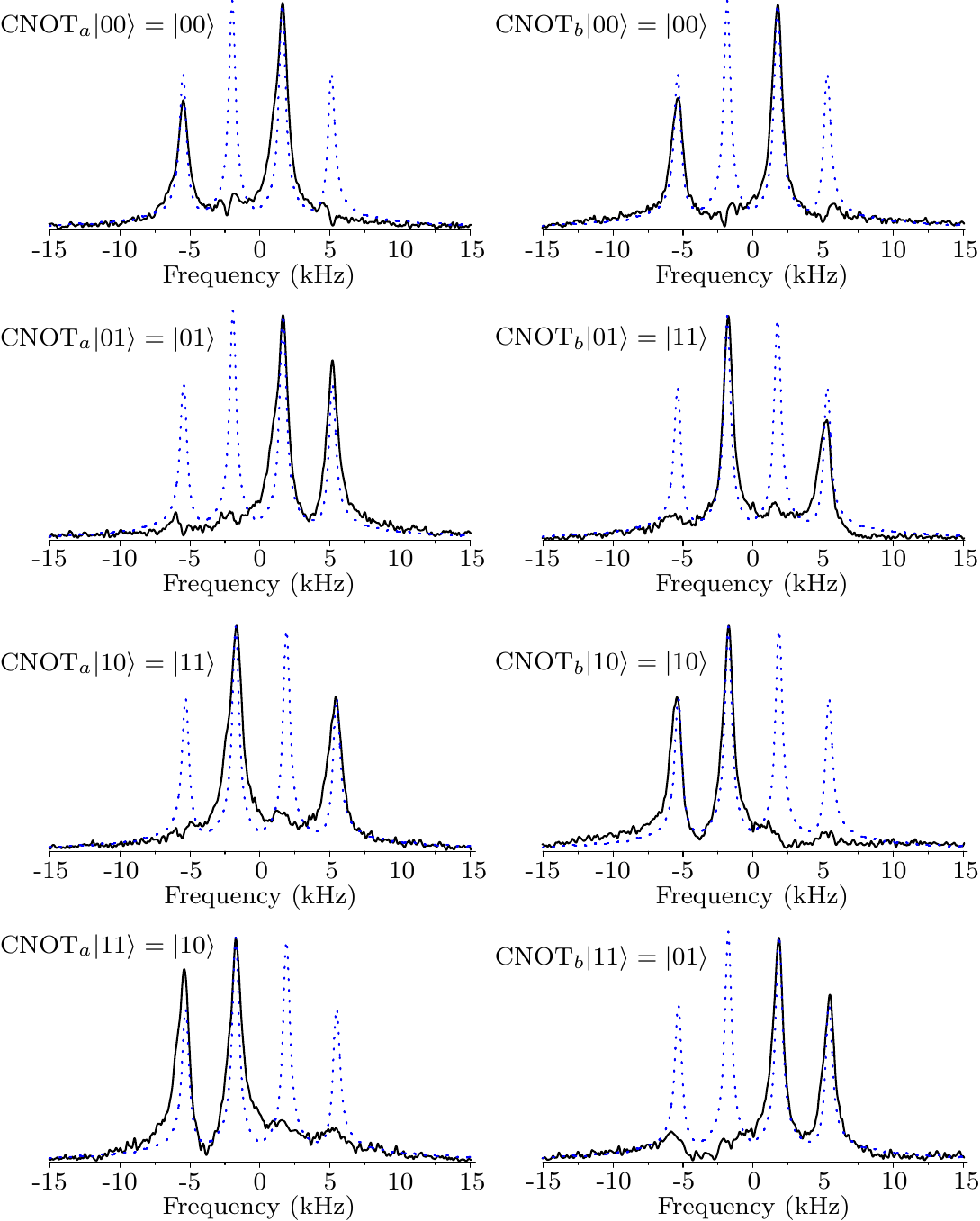}
\caption{$^{35}$Cl spectra of the CNOT gates on each PPS state after a $\pi/2$ pulse. The dashed lines correspond to the equilibrium spectrum for comparison.}
\end{figure}

We also created entangled pseudo-pure states by applying the Hadamard gate on each PPS state of the computational basis. The results shown in Fig.~8(a) correspond to the modulus of the spectra obtained after a $\pi/2$ reading pulse. The black dots are the theoretical amplitudes and the horizontal bars represent the standard deviation of the experimental amplitudes relative to the theoretical ones. We can see a good agreement between the theoretical and experimental data. Only the third spectral line of the H$_{ab}$PPS$_{10}$ and the fourth line of the H$_{ab}$PPS$_{11}$ presented a strong deviation from the expected values. For this reason, the fourth line of the H$_{ab}$PPS$_{11}$ was not included in the calculation of the amplitudes standard deviation represented by the horizontal bars. With that exception, the largest standard deviation was 12\% due to the H$_{ab}$PPS$_{00}$. In Fig.~8(b) is shown a second successive application of the Hadamard gate which, since it is a self-adjoint gate, must result in the original state. We can see a very good correlation with the amplitudes of the corresponding states in Eqs.~(\ref{eq34}) to (\ref{eq37}).

The deviations from the expected results can be mainly attributed to the RF amplitude inhomogeneity, since, in order to reach a good filling factor, the sample extended itself until the edges of the RF coil. NQR is also very sensitive to the sample temperature, where the quadrupole transition frequencies variations can be of the order of kHz per kelvin \cite{utton1967}. Therefore, small variations in the sample temperature are responsible for off-resonance pulse imperfections. In order to minimize such effects the pulse sequences were implemented with a large number of dummy scans and a relatively small number of scans. In this way, the thermal equilibrium in the presence of the RF radiation was established as good as possible before effective signal acquisition. Other source of error is, in a less extent, in the determination of the $\theta$ angle between the static magnetic field and the EFG symmetry axis.

The best way to evaluate the experimental performance of a quantum operation would be to implement quantum state or quantum process tomography \cite{chuang1998b,bonk2004,kampermann2005}. We are already developing such a method, which will be presented in a forthcoming article. With the $\pi/2$ rotation operation we have only partial access to the 2 q-bit system density matrix. Therefore, states that present spectral amplitudes very close to the theoretical expectations, as those presented in Fig.~5(b), can have, in fact, undesired coherences which do not appear in such spectra, but can be unveiled by a Hadamard gate operation, which mix together many different coherences. That could be an explanation for the strong deviation in the fourth spectral line of the H$_{ab}$PPS$_{11}$ state in Fig.~8(a).

\begin{figure}[h!]
\centering
\includegraphics[scale=0.8]{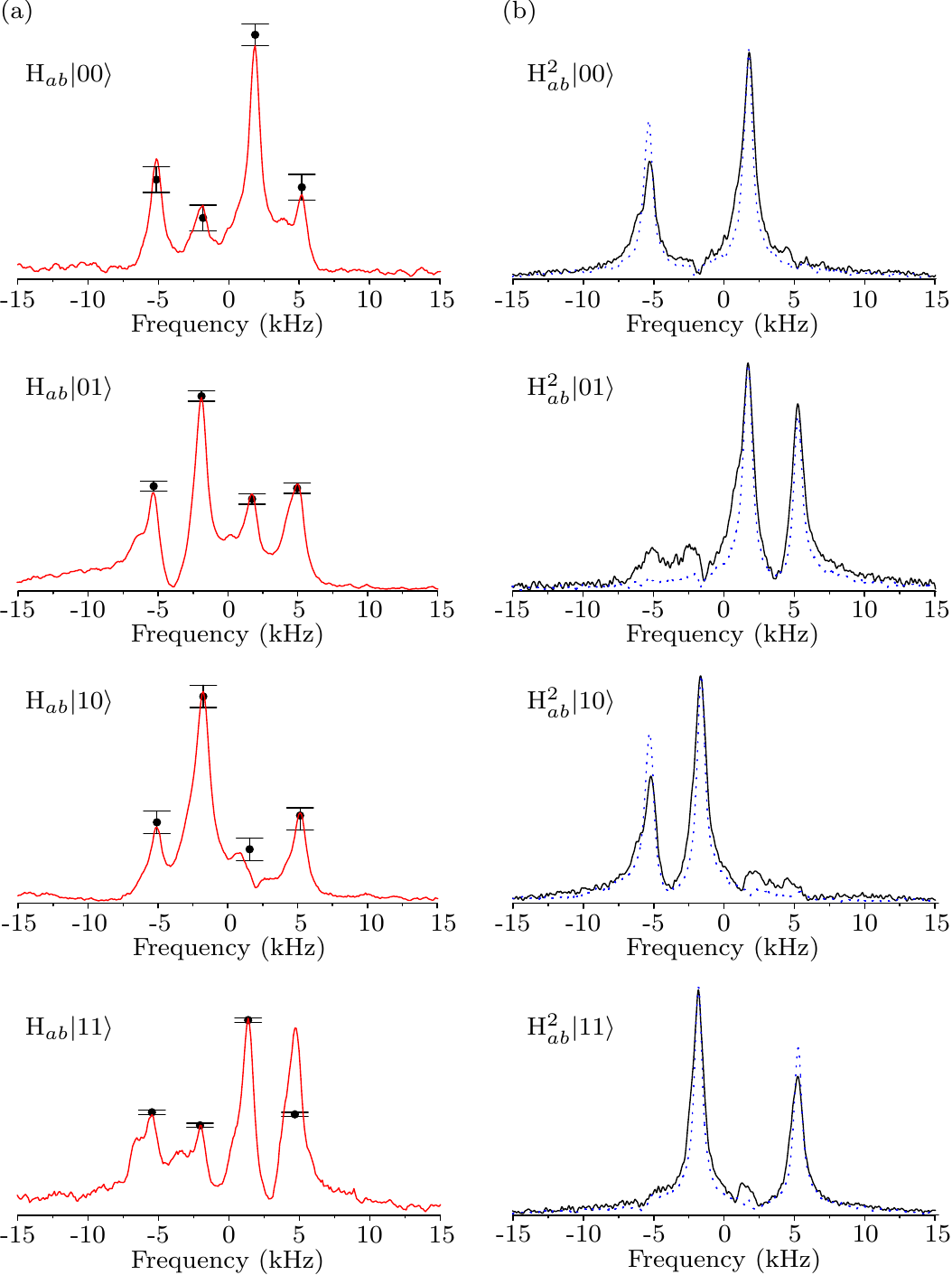}
\caption{$^{35}$Cl spectra of the Hadamard gate on both q-bits of each PPS state after a $\pi/2$ pulse. (a) One gate application. The black dots represent the theoretical amplitudes and the horizontal bars represent the standard deviation relatively to the experimental amplitudes. Only the fourth spectral line of the H$_{ab}|11\rangle$ state was not included in the standard deviation due to its strong disagreement. (b) Successive application of two Hadamard gates. The dashed lines correspond to the states before applying this operation.}
\end{figure}

\section{Conclusion}

The results presented in this work show that the nuclear spin quantum states can be finely controlled and read in a Quantum Information NQR context, showing results as good as solid state NMR. That was proved by the experimental implementation of the four pseudopure states associated to the spin 3/2 of the chlorine nuclei in the sodium chlorate single crystal. In our approach, a small static magnetic field and a linearly polarized RF field perturbation were applied in order to implement CNOT and Hadamard gates, whose results are in accordance to the expected ones. This indicates that coherence control and the quantum transitions available for the Zeeman perturbed NQR technique can be used to implement basic QIP protocols. Besides its performance similar to NMR, NQR presents other experimental physical parameters that can enhance studies involving the decoherence and relaxation of quantum states \cite{auccaise2008,soarespinto2011}. An actual subject of research is the study of quantum correlations at room temperature, specifically the determination of quantum discord in NMR, which could also be studied in the NQR context \cite{auccaise2011,maziero2013}. Many recent proposals of QIP in NMR quadrupole systems, such as the study of coherent spin states \cite{estrada2013} and bifurcation \cite{araujo2013}, could also be studied by NQR.

In order to full explore the quantum states evolution of low dimension Hilbert spaces using NQR it will be interesting to develop a Quantum State Tomography procedure. Future improvements include the design of more robust pulse sequences against pulse imperfections \cite{khaneja2005} or the design of coils with more homogeneous fields. A stable temperature control with sensibility on the order of milikelvin would also avoid off-resonance RF pulse imperfections. 

\section*{Acknowledgements}

This work was supported by Brazilian agencies FAPESP (2012/02208-5) and CNPq (483109/2011-8).

The authors also acknowledge Aparecido Fernandes de Amorim and Elderson C\'{a}ssio Domenicucci by the technical support.





\bibliographystyle{elsarticle-num}
\bibliography{NQR_QIP}







\end{document}